\documentclass{mn2e}
\usepackage{psfig}
\newcommand{\beq}{\begin{equation}}
\newcommand{\eeq}{\end{equation}}
\newcommand{\bea}{\begin{eqnarray}}
\newcommand{\eea}{\end{eqnarray}}
\begin{document}

\title[Jitter radiation from a relativistic blastwave] {Jitter
radiation images, spectra, and light curves from a relativistic
spherical blastwave}

\author[Morsony et al.]{Brian J. Morsony$^1$, Jared C. Workman $^1$,
Davide Lazzati$^1$ and Mikhail V. Medvedev$^{2,3}$ \\
$^1$ JILA, University of Colorado, 440 UCB, Boulder, CO 80309-0440, USA \\
$^2$ Department of Physics and Astronomy, University of Kansas, 
Lawrence, KS 66045 \\
$^3$ Institute for Nuclear Fusion, RRC ``Kurchatov Institute'', Moscow 123182, Russia}

\maketitle

\begin{abstract}
We consider radiation emitted by the jitter mechanism in a
Blandford-McKee self-similar blastwave. We assume the magnetic field
configuration throughout the whole blastwave meets the condition for
the emission of jitter radiation and we compute the ensuing images,
light curves and spectra. The calculations are performed for both a
uniform and a wind environment. We compare our jitter results to
synchrotron results. We show that jitter radiation produces slightly
different spectra than synchrotron, in particular between the
self-absorption and the peak frequency, where the jitter spectrum is
flat, while the synchrotron spectrum grows as $\nu^{1/3}$. The
spectral difference is reflected in the early decay slope of the light
curves. We conclude that jitter and synchrotron afterglows can be
distinguished from each other with good quality
observations. However, it is unlikely that the difference can explain
the peculiar behavior of several recent observations, such as flat
X-ray slopes and uncorrelated optical and X-ray behavior.
\end{abstract}


\section{Introduction}

The synchrotron external shock model (Meszaros \& Rees 1997) has been
used with great success to model GRB afterglows since their discovery
(Wijers, Rees \& Meszaros 1997; Piran 1999; Panaitescu \& Kumar 2001;
Zhang \& Meszaros 2004). However, its physical foundations have been
put under debate, principally because a coherent equipartition
magnetic field is assumed to be generated at the shock front. In
addition, the fraction of internal energy that is given to the
magnetic field is assumed independent of the location in the blastwave
(the distance from the shock) and of the properties of the shock
(e.g. its Lorentz factor).

The two latter assumptions have been tested phenomenologically on
afterglow data. Rossi \& Rees (2003) considered a magnetic field that
is allowed to decay behind the shock front. They concluded that
afterglow observations require a magnetic field that has an
approximately uniform intensity throughout the blastwave (or at least
within a distance $R/\Gamma^2$ from the shock front). Yost et
al. (2003) studied the possibility that the fraction of the internal
energy stored in the magnetic field component grows with the decrease
of the Lorentz factor of the fireball. They concluded that this effect
could explain a sample of radio afterglows that have flat late time
radio decay.

Even though these works present interesting constraints on the
behavior of the magnetic field, they do not address the mechanism that
generates it, its structure, and whether the radiation mechanism is
really synchrotron or not. Analytic (Medvedev \& Loeb 1999) and
numerical work (Silva et al. 2003; Nishikawa et al. 2003; Frederiksen
et al. 2004; Medvedev et al. 2005; Spitkovsky 2008; Chang, Spitkovsky
\& Arons 2008) have started to unveil recently the mechanism by which
a magnetic field component is generated behind a relativistic
collisionless shock front. All the studies find that the magnetic
field is generated through some kind of two-stream instability, such
as the Weibel instability (Weibel 1959). Magnetic fields are generated
behind the shock with an energy of the order of 30 per cent of the
total available internal energy. They are observed to subsequently
decay and, possibly, to stabilize at a level of a fraction of a per
cent of the internal energy. Whether they decay even further deep into
the shocked material is a matter of debate. The structure of the field
is observed to be highly tangled and planar behind the shock front and
to grow more coherent deep into the blastwave.

The correlation length of the magnetic field right behind the shock is
so short that the synchrotron approximations do not hold and a new
radiative regime sets in. This form of radiation, called jitter
radiation (Medvedev \& Loeb, 1999; Medvedev 2000, 2006 or, sometimes,
diffusive synchrotron radiation, Fleishman 2006ab), has been studied
analytically, semi-analytically and numerically (Medvedev \& Loeb
1999; Medvedev 2000; Frederiksen et al. 2004; Medvedev 2006; Fleishman 2006a;
Medvedev et al. 2007; Workman et al. 2008). In this work we apply the
theory of jitter radiation to compute the emission from a Blandford \&
McKee (1976) blastwave, taking into account its structure, the special
relativistic effects, and the light propagation time. The results
include spectra, lightcurves and images, analogously to what is done by
Granot, Piran \& Sari (1999a) for the synchrotron case.

This paper is organized as follows: In \S~2 we describe the numerical
methods applied to integrate the radiation spectrum over the BM
solution and over surfaces of equal arrival time; in \S~3 we describe
our results and in \S~4 we summarize and discuss them.

\section{Modeling The Afterglow}

In this section we describe the process by which we solve for the
physical shape, evolution, and internal properties of a Gamma-Ray
Burst (GRB) blast wave.  We loosely follow the outline used by Granot
et al.  (1999a) but we expand on their work by allowing for both a
constant density ISM and a wind environment such as would possibly be
found near the progenitor stars of long duration Gamma Ray Bursts.  We
will describe the physical setup of the problem and then describe the
method used to compute images, spectra, and light curves assuming that
the whole blastwave radiates jitter radiation. See Medvedev \& Loeb
(1999), Medvedev (2000) and Workman et al. (2008) for a thorough discussion of the
conditions under which jitter radiation is produced.

\subsection{The Blandford McKee Solution} 

To calculate the spectrum of a GRB afterglow emitting jitter radiation
we first have to compute the local physical quantities in the
blastwave and the time evolution of the afterglow in the relevant
geometry of the external medium. As soon as the external shock has
swept up a mass comparable to the rest mass of the fireball over its
Lorentz factor, the blastwave enters the self-similar stage (Blandford
\& McKee 1976).

Following Blandford \& McKee (1976) we define a similarity variable
as:
\beq \chi(r) = 1 + 4(4-k)\gamma^2_{f}(1-\frac{r}{R}) \eeq
where $R$ is the radius of the shock, $r$ is the distance to the
center of the shock, $\gamma_{f}$ is the Lorentz factor of the
material immediately behind the shock front and $k$ describes the
properties of the external medium. For a constant density profile,
$k=0$, while for a wind environment $k=2$.  Using this similarity
variable, we can solve for the comoving number density ($n^\prime$),
the comoving energy density ($e^\prime$), and the Lorentz factor
($\gamma$) of the material anywhere within the expanding volume with
the following equations:
\beq
\label{nprime}
n' = 4n\gamma_f\chi^{\frac{1}{2}}\chi^{\frac{(2k-7)}{(4-k)}},
\eeq
\noindent
where $n(r)=n_{ISM}$ for the constant density case and
$n(r)=\frac{Ar^{-2}}{m_p}$ for the wind environment with
$A=\frac{\dot{M}_w}{4\pi V_w}$ g/cm,
\beq
\label{eprime}
e' = 4nm_pc^2\gamma_f^2\chi^{\frac{(4k-17)}{(12-3k)}},
\eeq 
\noindent
and
\beq
\label{gam}
\gamma=\gamma_f\chi^{\frac{-1}{2}}.
\eeq

\noindent
To solve for the above quantities, we need to find $\gamma_f$, the
Lorentz factor of the material directly behind the shock front that
defines the blastwave outermost edge. It is given by (e.g., Chevalier
\& Li, 2000) 
\beq \gamma_{f}=\left[\frac{(17-4k)E}
{16\pi m_pn(R)R^{3}c^2}\right]^{\frac{1}{2}}.  
\eeq
\noindent
where $E$ is the energy of the blastwave.

\begin{figure*}
\psfig{file=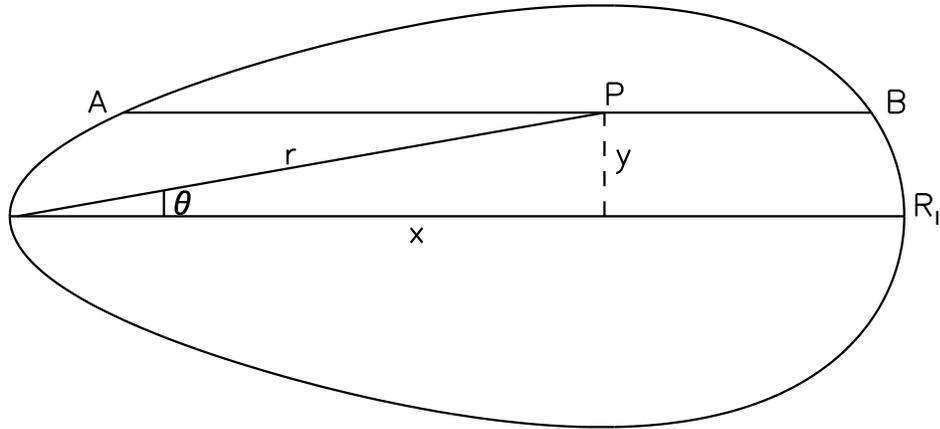}
\caption{{Schematic diagram of the region from which photons reach a
distant observer simultaneously (Granot, Piran \& Sari, 1999a).  The
line A-B represents a line of sight along which the radiative
transfer equation is solved to find $I_{\omega}(y)$. }
\label{fig:1}}
\end{figure*}

Let us consider now the point $P$ in Fig.~\ref{fig:1}. It is
characterized by a distance $r$ from the center of the explosion and
an angle $\theta$ with respect to the line of sight. 
Velocity at this point is in the $\hat{r}$ direction.
The point $P$ begins to emit radiation when it is crossed 
by the forward shock at the observed time:
\begin{equation}
T_{obs}=\frac{r}{c}
\left(1-\cos(\theta)+\frac{1}{4(4-k)\gamma_f^2}\right)
\end{equation}
The radiation observed at time $T_{obs}$ is obtained therefore by
integration over the locations $(r,\theta)$ that satisfy:
\begin{equation}
\frac{r}{c} \left(1-\cos(\theta)+\frac{1}{4(4-k)\gamma_f^2}\right)
\le T_{obs}
\label{eq:condition}
\end{equation}

At any location that satisfies Eq.~\ref{eq:condition}, the ratio $R/r$
that is needed to compute the self-similar variable $\chi$ is computed
through the equation:
\begin{equation}
T_{obs}=\frac{R}{c}
\left(1+\frac{1}{4(4-k)\gamma_f^2}\right)-\frac{r}{c}\cos(\theta)
\label{cr}
\end{equation}
where $R$ is the radius of the shock at the laboratory time at which
the photon was emitted at radius $r$ to be observed at infinity at
time $T_{obs}$.

The energy density of the magnetic field and of non-relativistic
particles is assumed to be a constant fraction of the local internal
energy, as derived from the BM blast wave solution. Such a
prescription is analogous to adiabatic evolution of the population of
relativistic electrons with the density of particles (Beloborodov
2005).  We use the full form for Weibel fields described in detail in
Medvedev (2006) and Workman et al. (2008) to describe the geometry and
correlation function of the magnetic field. We assume that the
properties of the magnetic field stay constant throughout the
afterglow and throughout the blast wave. We parametrize the magnitude
of the magnetic field according to the usual prescription
\beq
\label{bfield}
\frac{\langle B \rangle^2}{8\pi}=\epsilon_{B}e',
\eeq
\noindent  
where $\epsilon_B$ represents the equipartional fraction of the
magnetic field with the energy density.

\subsection{The Observed Jitter Spectrum}

To determine $P'_{\rm tot}(\omega)$, the total power emitted per unit
frequency by an ensemble of electrons radiating in a unit volume, we
perform the following integral
\beq
P'_{\rm tot}(\omega)=\int_{\gamma_{min}}^\infty n'(\gamma)p'(\omega)d\gamma.
\label{pensemble}
\eeq
\noindent
The number density $n$ and the minimum Lorentz factor ($\gamma_{min}$)
in the integral are derived by basic assumptions about the total
number and energy of electrons, the interested reader is referred to
Workman et al. (2008).  Once the above quantity has been computed we
assume isotropic radiation and, following Rybicki and Lightman (1979),
we define the emission coefficient, $j'_{\omega}$ as
\beq
j'_{\omega}=\frac{P'_{\rm tot}(\omega)}{4 \pi}.  
\eeq
\noindent
Following the process outlined in Medvedev et al. (2007), Workman et al. (2008); Granot,
Piran, and Sari (1999b), and Rybicki and Lightman (1979) we define the
absorption coefficient as
\beq 
\alpha'_{\omega}= \frac{(p+2)\pi^2 k}
{m_e \omega^2}\int_{\gamma_{min}}^\infty
\gamma^{-(p+1)}P'(\omega)d\gamma.
\label{absorb}
\eeq
\noindent
Once we have found the necessary coefficients we use the fact that
$\alpha_{\omega}\omega$ and $\frac{j_{\omega}}{\omega^2}$ are both
Lorentz invariants with
$\omega=\omega'\gamma_b(1+\frac{v}{c}\cos\theta')$; where $\gamma_{b}$
and $v$ refer to the bulk Lorentz factor and velocity of the afterglow
and $\cos\theta'$ is defined by the angle between a photon's path and
the co-moving volume's direction. Using this property we can re-write
the differential equation for specific intensity in our frame, given
by
\beq
\frac{dI_{\omega}}{ds} = j_{\omega}-\alpha_{\omega} I_{\omega}
\eeq
\noindent
directly in terms of comoving quantities.  We assume that there is no
scattering or frequency redistribution (see Fleishman 2006ab).  This
allows us to calculate the spectrum by numerically solving for a
series of one dimensional, line of sight, integrals.  We define our
integration region as the three dimensional volume of constant arrival
time, which is given by equation \ref{cr}.  As the jitter radiation
depends on the angle of emission in the co-moving frame, we solve for
$\theta'$ using $\theta$ and $\beta$ by inverting the equation
\beq
\label{thetainv}
cos(\theta)=\frac{cos(\theta')+\beta}{1+\beta cos(\theta')}
\eeq
Once we have found the comoving angle at each point, we use the the
Blandford-McKee solution to find the values of $\gamma$, $n'$, $e'$,
and the magnetic field at each point and solve for $j'_{\omega}$ and
$\alpha_{\omega}'$.

With these functions known throughout the volume of constant arrival
time, we solve the equation for observed specific intensity, at a
given frequency, along lines of sight that span the vertical extent of
the surface, giving us $I_{\omega}(y)$ where y is the perpendicular
distance from the center of the source (see Fig.~\ref{fig:1}).

Cooling is added analytically by changing the spectral slope to 
$\omega^{p-2}$ above a frequency of 
\beq
\omega_c = \frac{8.5\times10^{16}}{2\pi} E_{52}^{-1/2} 
(\frac{\epsilon_B}{0.001})^{-3/2} n_{ISM}^{-1} t_{days}^{-1/2}
\eeq
(from Sari et al. 1998) for the ISM case and 
\beq
\omega_c = \frac{5.7\times10^{15}}{2\pi} E_{52}^{-1/2} 
(\frac{\epsilon_B}{0.001})^{-3/2} A_{*}^{-2} t_{days}^{1/2}
\eeq
(from Chavalier \& Li 2000) for the wind case, 
where $A_{*} = \frac{A}{3\times10^{35}~cm^{-1}}$.  
However, for the parameters and times considered in this paper,
cooling only becomes important above $32$~keV for the ISM case and
$24$~keV for the wind case.

Once we have solved for the specific intensity in the observer frame,
the observed flux at the detector is given by
\beq
F_{\omega}=2\pi\frac{1}{d^2}\int_0^{R_{\perp,max}} I_{\omega}(y)ydy,
\eeq 
\noindent
where $d$ is the distance to the afterglow and $R_{\perp,max}$ is the
largest perpendicular extent of the afterglow (as in Granot et
al. 1999a).

\section{Results}

In order to compare synchrotron and jitter radiation we calculate the
spectrum and time evolution for both radiation mechanisms for the case
of a constant (ISM) and a wind external medium, for a total of 4
models.  Analytical lightcurves for these models are derived in Medvedev et al. (2007).
In all four cases the total isotropic equivalent energy in
the blast wave was $E=10^{53}$~erg, the magnetic field equipartition
fraction was $\epsilon_B=10^{-5}$, and the fraction of internal energy
in electrons was $\epsilon_e=0.01$.  For the ISM case the external
density was $n=1$ while for the wind medium we assumed
$\dot{M}=10^{-6}M_{\sun}$/yr and $v_w = 1000$~km/s giving $A_{*}=0.1$.

\begin{figure*}
\psfig{file=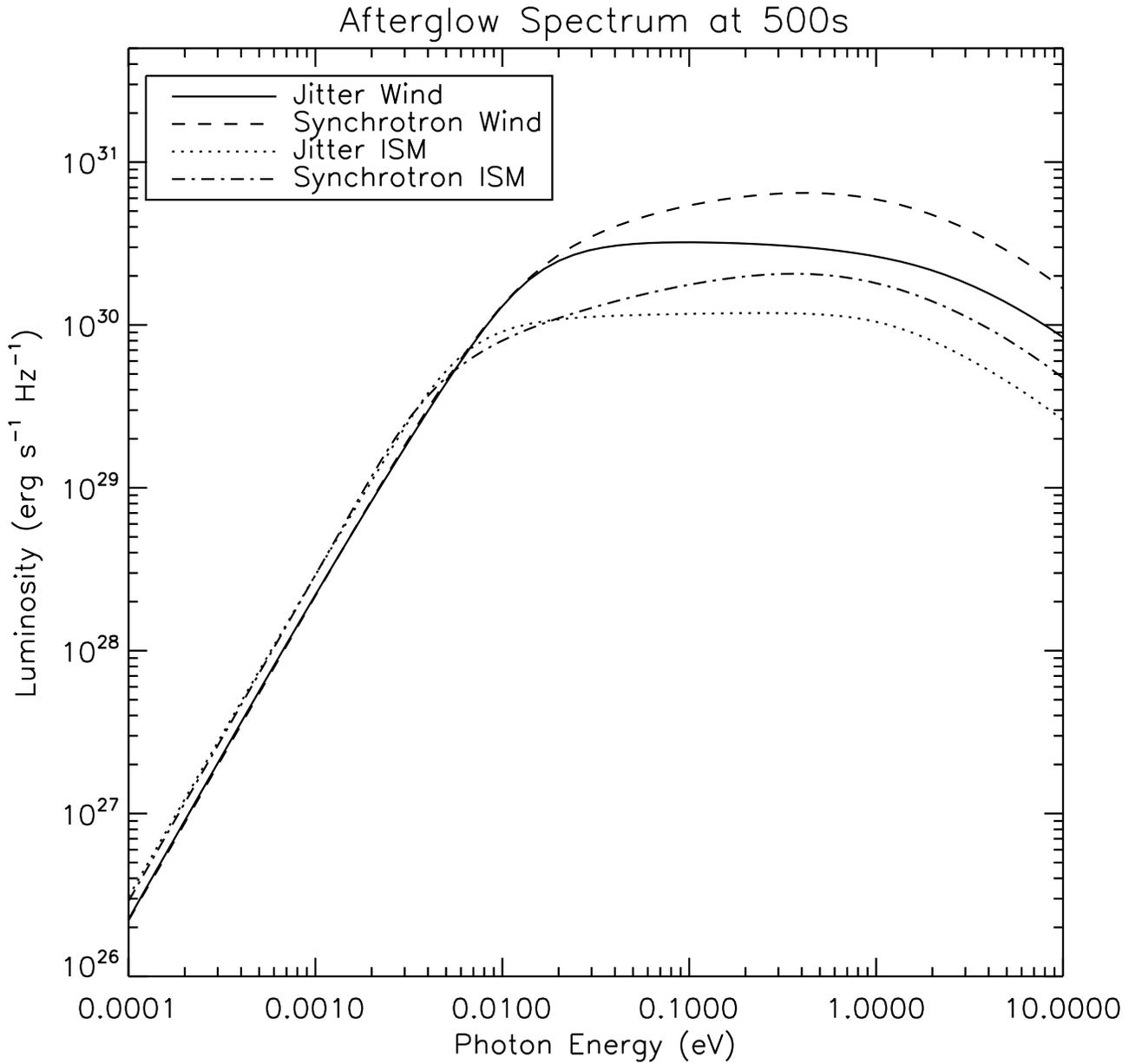}
\caption{{Comparison of afterglow spectra at $t=500$ seconds.  Spectra
plotted are for radiation from a shock expanding into a wind or
ISM-type medium produced by jitter and synchrotron radiation.  Both
jitter spectra are flat between the peak frequency and optically thick
cutoff, while the synchrotron spectra increase as $\omega^{1/3}$
before reaching a broad peak.  The wind cases are optically thick at
lower frequencies. }
\label{fig:2}}
\end{figure*}

\begin{figure*}
\psfig{file=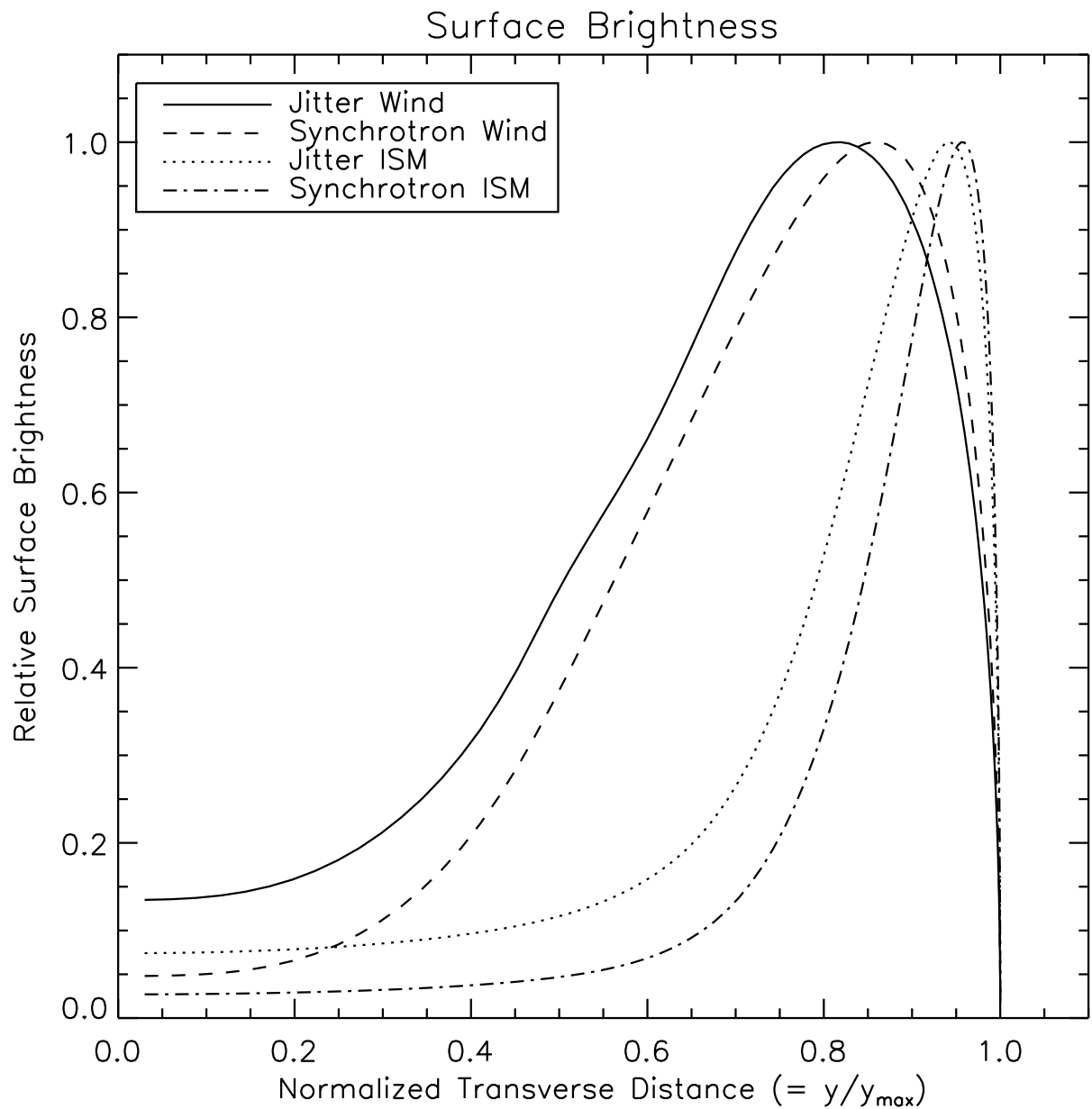}
\caption{{Comparison of surface brightness at $1$~eV at $t=1000$ seconds 
for the same 4 models of Fig.~\ref{fig:2}.  In all cases, the emission
is concentrated towards the outer edge.  Normalized transverse distance is defined as $y/y_{max}$ with $y$ defined as in Fig.~\ref{fig:1}.}
\label{fig:3}}
\end{figure*}

A spectrum for each of these 4 models at $t=500$s is plotted in
Fig.~\ref{fig:2}.  At this time, the two synchrotron models have a
peak frequency of about $0.3$~eV.  At lower frequencies the spectrum
increases as $\omega^{1/3}$ and at higher frequencies it decreases as
$\omega^{-(p-1)/2}$.  The self-absorption break is observed at
$\sim10^{-2}$~eV.  The two jitter radiation models have similar
behavior to the synchrotron models above the peak frequency and in the
optically thick regime.  However, the jitter models have a nearly flat
spectrum below the peak frequency rather than a $\nu^{1/3}$ branch.
Even though jitter radiation can produce a spectral slope as steep as
$\omega^{1}$ at small angles, the integrated emission is dominated by
material at large comoving angles (see Fig.~\ref{fig:3}) which produce
a flat spectrum for jitter radiation (see Medvedev 2006; Medvedev et
al. 2007; Workman et. al. 2008).  A good quality spectrum over a wide
frequency range can distinguish between the synchrotron and jitter
radiation mechanisms if the $\nu^{1/3}$ or flat portion of the
spectrum is captured. However, the spectrum alone cannot separate ISM
and wind external media.

\begin{figure*}
\psfig{file=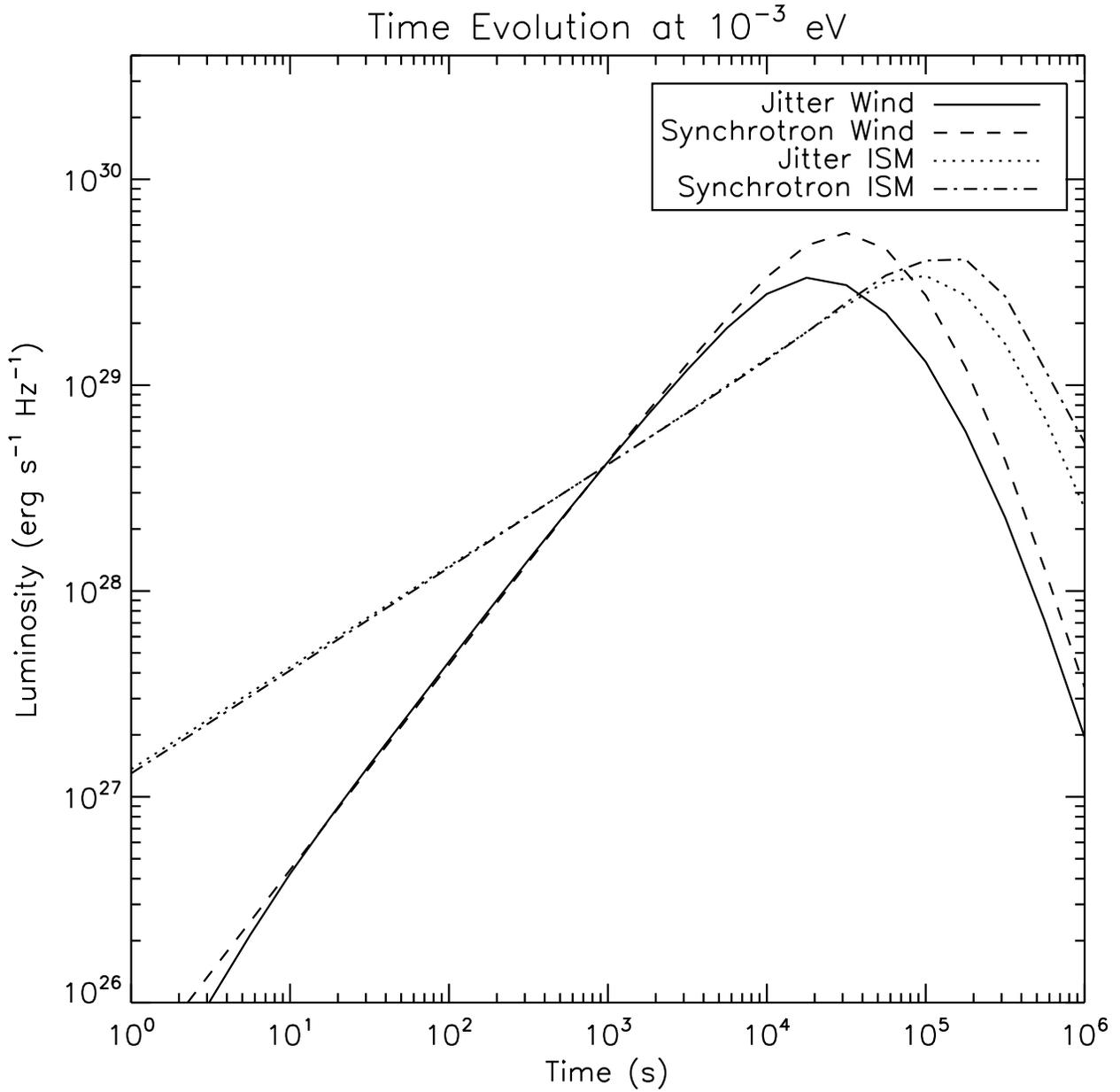}
\caption{{Simulated light curves at $h\nu=10^{-3}$~eV 
($\lambda=1.24$~mm) for the 4 models of Fig.~\ref{fig:2}.  All 4
models are optically thick at this frequency.  The ISM models decrease
in luminosity as $t^{1/2}$, while the wind models increase as
$t^{1}$.}
\label{fig:5}}
\end{figure*}

At very low frequency (see Fig.~\ref{fig:5}) all models are initially
optically thick.  This results in an increase of luminosity as
$t^{1/2}$ for the two ISM models and an increase as $t^{1}$ for the
two wind models.  The radiation mechanism does not significantly
affect the results in the optically thick regime.

\begin{figure*}
\psfig{file=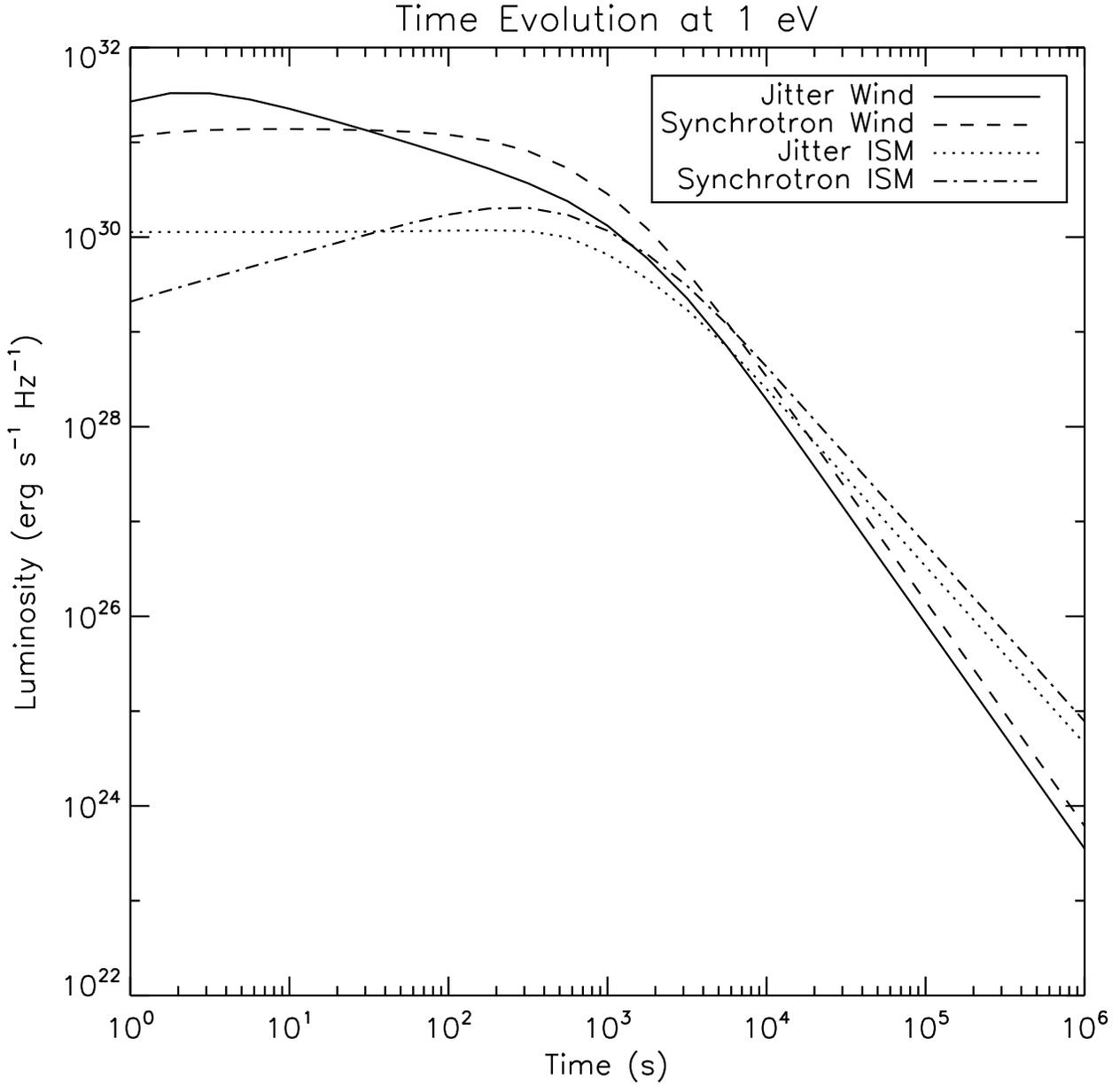}
\caption{{Simulated light curve at $h\nu=1$~eV ($\lambda=1240$~nm) 
for the 4 models of Fig.~\ref{fig:2}.  The synchrotron-ISM model
luminosity increases at early times as $t^{1/2}$.  The
synchrotron-wind and jitter-ISM models are both initially flat before
transitioning to a power-law decay as the peak frequency decreases
below $h\nu=1$~eV.  The jitter-wind model initially decreases as
$t^{-1/2}$ before transitioning to a steeper power-law decay.  }
\label{fig:6}}
\end{figure*}

At higher frequency all four models are optically thin.  A comparison
of the light curves for the four models at $1$~eV is shown in
Fig.~\ref{fig:6}. For the ISM case, the synchrotron spectrum initially
increases as $t^{1/2}$ and then transitions to a steep decay
($t^{-3(p-1)/4}$) as the peak frequency of the spectrum decreases
below $1$~eV.  This light curve is similar to the low frequency
synchrotron-ISM and jitter-ISM models is Fig.~\ref{fig:5}.  With a
wind external medium and synchrotron radiation, the light curve is
initially flat before transitioning to a steep decay
($t^{-(3p-1)/4}$).  If the jitter radiation mechanism is used, the
slope of the light curves are initially decreased by a factor of
$t^{-1/2}$, giving a flat initial slope for the jitter-ISM model and a
decrease of $t^{-1/2}$ for the jitter-wind case.  The result is that
in this regime the jitter-ISM and synchrotron-wind models are not
distinguishable from each other from the light curve alone, as they
both begin with a flat intensity.

\begin{figure*}
\psfig{file=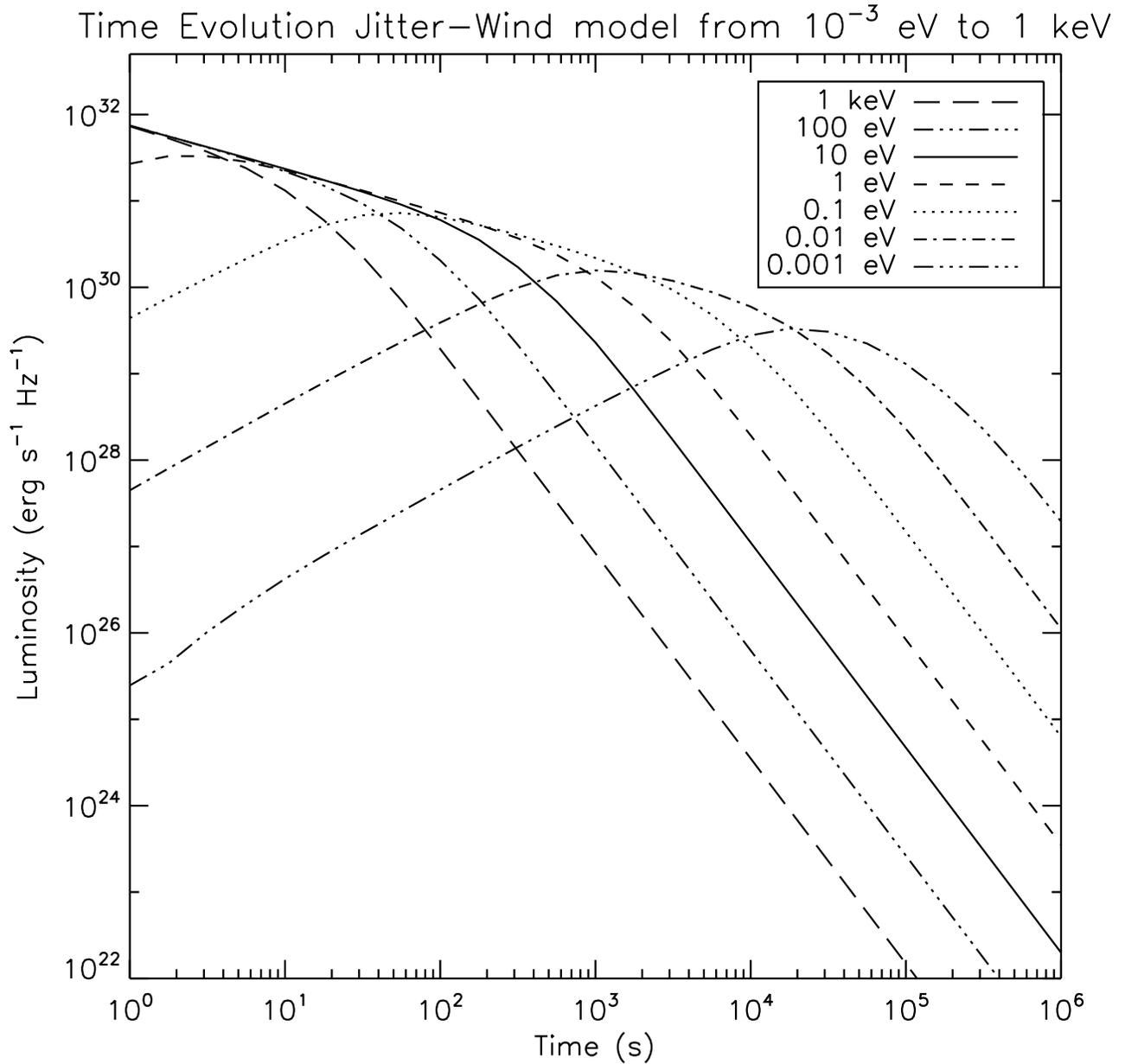}
\caption{{Comparison of light curves for jitter-wind model at
different energies.  The transition time between the luminosity
scaling as $t^{-1/2}$ and $t^{-(3p-1)/4}$ is dependent on frequency
and gives the appearance of a chromatic break in a power-law decay
light curve.  The time of this break ranges from about $t=10$s to
$t=10^5$s for the frequencies shown in the figure.}
\label{fig:7}}
\end{figure*}

For the jitter-wind case, the luminosity initially decreases as
$t^{-1/2}$.  This gives the light curve the appearance of a broken
powerlaw decay.  The break occurs at about $10^3$s at $1$~eV for this
example, but the time of the break depends on frequency as $t_{break}
\propto \omega^{-2/3}$.  Fig.~\ref{fig:7} shows light curves from the
jitter-wind model for energies between $10^{-3}$~eV and $1$~keV.  The
jitter wind model naturally gives rise to a power law decay with a
chromatic break between slopes of $t^{-1/2}$ and $t^{-(3p-1)/4}$.  At
lower frequencies in Fig.~\ref{fig:7}, the initial slope is increasing
as the ejecta are initially optically thick.  Observations of the
jitter-wind model would show a chromatic break over a range of
frequencies or a steep decay at high frequency with a shallow decay
and then a break at lower frequencies, depending on the time range and
frequencies of the observations.

\begin{figure*}
\psfig{file=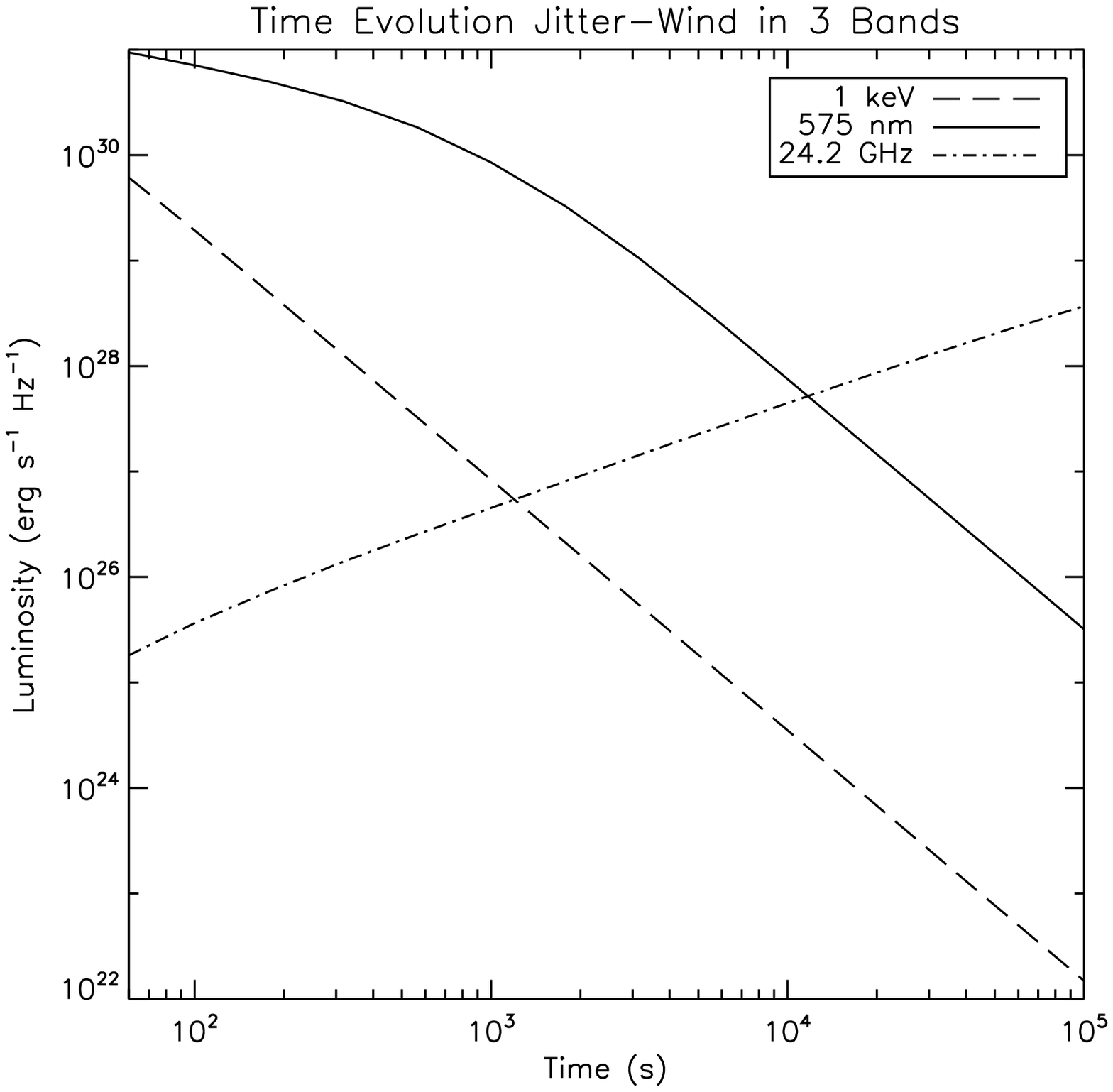}
\caption{{Comparison of light curves for jitter-wind model in the 
X-ray ($h\nu=1$~keV), visible ($\lambda=575$~nm) and radio
($\nu=24.2$~GHz) bands, observed from $t=60$s to $t=10^5$s.  The x-ray
light curve is a single powerlaw decay, the visible is a broken
powerlaw and the radio, in the optically thick regime, is a powerlaw
increase.}
\label{fig:9}}
\end{figure*}

Fig.~\ref{fig:9} shows light curves from the jitter-wind model in the
x-ray ($1$~keV), visible ($575$~nm), and radio ($24.2$~GHz).  In this
case, the x-ray observations show a single powerlaw decay, the visible
shows a broken powerlaw, and the radio shows a powerlaw increase.

\section{Discussion and conclusions}

We have calculated the spectra and light curves produced by jitter
radiation from a relativistic fireball expanding into either an ISM
(constant density) or wind ($\rho \propto r^{-2}$) external medium
including self absorption and electron cooling, and then compared
these results to results for synchrotron radiation in identical
cases. We improved upon our previous computations (Medvedev et
al. 2007) by performing the integration over the blastwave and taking
into account the light travel time of the photons.  

The primary difference between jitter and synchrotron radiation is
that jitter radiation produces a flat spectral slope below the peak
frequency, rather than the $\omega^{1/3}$ slope seen in synchrotron
radiation.  Jitter and synchrotron radiation are not distinguishable
in the optically thick regime or above the peak frequency, only in
between.  The flat spectral slope of jitter radiation has the effect
of decreasing the slope of the light curve by a factor of $t^{-1/2}$
compared to synchrotron radiation in the relevant frequency and
temporal intervals.  This means that the slope of the light curve,
while the observed frequency is below the peak frequency, decreases
from $t^{1/2}$ for synchrotron to $t^{0}$ for jitter in the ISM case
and decreases from $t^{0}$ to $t^{-1/2}$ for the wind case.  This
indicates that the jitter-ISM and synchrotron-wind cases would not be
distinguishable from a single light curve as they both have a $t^{0}$
slope.

In the last four years, {\it Swift} observations, combined with
ground-based optical and radio telescopes have put the standard
synchrotron external shock model under scrutiny. The new observations
have shown temporally flat X-ray light curves (Nousek et al. 2006),
chromatic breaks (e.g., Huang et al. 2007), and uncorrelated
multi-band behavior (e.g., Blustin et al. 2006; Dai et al. 2007). The
light curves and spectra of jitter and synchrotron radiation are
different. However, the differences appear to be too subtle to explain
the atypical afterglow observations summarized above. Even though
jitter radiation can produce flat lightcurves, it does so in all
wavebands and not only in the X-rays, and cannot explain the steep
decay observed, for example, in GRB 070110 (Troja et
al. 2007). Chromatic breaks can be produced by jitter radiation as well
as by synchrotron, the only difference being the amount of change in
the pre- and post-break decays. Pending a more quantitative comparison
of the jitter light curves and spectra with the data, it seems that
jitter is a viable radiation mechanism to explain the behavior of GRB
afterglows. It shares, however, all the problems of synchrotron in the
explanation of some of the most recent data. Some possible explanation
and a review of the difficulties can be found, for example, in
Ghisellini et al. (2007); Liang, Zhang \& Zhang (2007); Uhm \&
Beloborodov (2007); Panaitescu (2008).


\section*{Acknowledgments}

This work was supported by NSF grants AST-0407040 (JW), AST-0307502
(BM, DL) and AST-0708213 (MM), NASA Astrophysical Theory Grants NNG04GL01G and NNX07AH08G
(JW) and NNG06GI06G (BM, DL), Swift Guest
Investigator Program grants NNX06AB69G (JW, BM, DL) and NNX07AJ50G and NNX08AL39G (MM), and DoE grants
DE-FG02-04ER54790 and DE-FG02-07ER54940 (MM).  MM gratefully acknowledges support from the
Institute for Advanced Study.


\end{document}